\documentclass[runningheads]{llncs}
\usepackage{graphicx}
\usepackage{amsmath}
\usepackage{multirow}
\usepackage{amssymb}
\usepackage{booktabs}
\usepackage{tikz,pgfplots}
\usetikzlibrary{spy}

\usepackage{textcomp}
\usepackage{gensymb}

\newcommand{\repeatthanks}{\textsuperscript{\thefootnote}}

\begin{document}
	\title{RinQ Fingerprinting: Recurrence-informed Quantile Networks for Magnetic Resonance Fingerprinting}
	\titlerunning{RinQ Fingerprinting}
	\author{Elisabeth~Hoppe\inst{1}\textsuperscript{,}\thanks{These authors have contributed equally and are listed in alphabetical order.}  
		\and
	Florian~Thamm\inst{1}\textsuperscript{,}\repeatthanks 
	\and
	Gregor~K{\"o}rzd{\"o}rfer\inst{2}
	\and
	Christopher~Syben\inst{1}
	\and
	Franziska~Schirrmacher\inst{1}
	\and
	Mathias~Nittka\inst{2}
	\and
	Josef~Pfeuffer\inst{2}
	\and
	Heiko~Meyer\inst{2}
	\and
	Andreas~Maier\inst{1}
	}

	\authorrunning{E.~Hoppe et al.}

	\institute{Pattern Recognition Lab, Department of Computer Science, Friedrich-Alexander-Universit{\"a}t Erlangen-N{\"u}rnberg, Erlangen, Germany   \email{\{elisabeth.hoppe,florian.thamm\}@fau.de} \\ \and
	MR Application Development, Siemens Healthcare, Erlangen, Germany}
	
	\maketitle

\begin{abstract}
	
	Recently, Magnetic Resonance Fingerprinting (MRF) was proposed as a quantitative imaging technique for the simultaneous acquisition of tissue parameters such as relaxation times $T_1$ and $T_2$. Although the acquisition is highly accelerated, the state-of-the-art reconstruction suffers from long computation times: Template matching methods are used to find the most similar signal to the measured one by comparing it to pre-simulated signals of possible parameter combinations in a discretized dictionary. Deep learning approaches can overcome this limitation, by providing the direct mapping from the measured signal to the underlying parameters by one forward pass through a network. In this work, we propose a Recurrent Neural Network (RNN) architecture in combination with a novel quantile layer. RNNs are well suited for the processing of time-dependent signals and the quantile layer helps to overcome the noisy outliers by considering the spatial neighbors of the signal. We evaluate our approach using in-vivo data from multiple brain slices and several volunteers, running various experiments. We show that the RNN approach with small patches of complex-valued input signals in combination with a quantile layer outperforms other architectures, e.g. previously proposed Convolutional Neural Networks for the MRF reconstruction reducing the error in $T_1$ and $T_2$ by more than 80\,\%. 
	
\end{abstract}
\section{Introduction}

\begin{figure}
	\centering
	\includegraphics[width=0.85\textwidth]{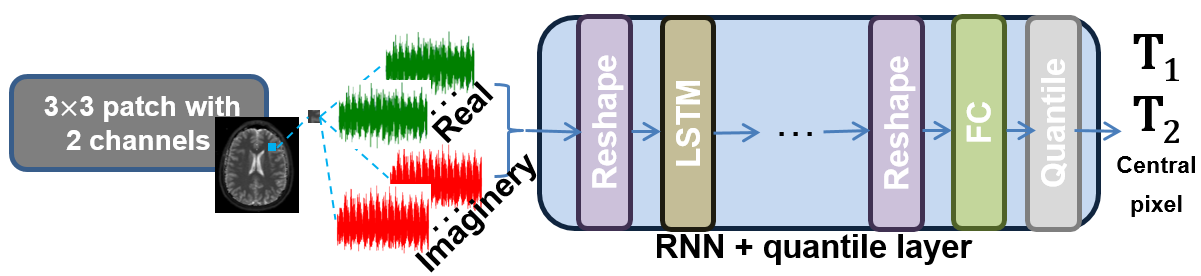}
	\caption{Overview over the MRF reconstruction process using deep learning. We map the reconstruction process using a Recurrent Neural Network with complex-valued input signals in combination with a quantile layer. LSTM: Long Short-Term Memory layer, FC: Fully Connected layer.} \label{fig:overview}
\end{figure}

One disadvantage of the most currently used Magnetic Resonance Imaging (MRI) techniques is the
qualitative nature of the images, thus in most cases no absolute values of the underlying physical tissue parameters, e.g. $T_1$ and $T_2$ relaxations, are obtained. Magnetic Resonance Fingerprinting (MRF) was recently proposed to overcome this limitation: It provides an accelerated acquisition of time signals which differ with the various tissue types by using randomly modified parameters during the acquisition (e.g. Flip Angle (FA) or Repetition Time (TR)) and strong undersampling with spiral readouts. These signals are compared to simulated signals of possible parameter combinations of $T_1$ and $T_2$ and quantitative maps are reconstructed~\cite{jiang2015mr,ma2013magnetic}. However, this state-of-the-art approach suffers from high computational effort: Every measured signal is compared to every simulated signal using template matching algorithms. Due to storage and computational limitations, this dictionary can only have a finite amount of possibilities and thus the maps are limited to these parameter combinations and can be erroneous~\cite{wang2014mrf}. The more combinations the dictionary contains, the more expensive is the reconstruction in terms of time and storage. In order to provide continuous predictions, to accelerate this process and to eliminate the burden of high storage requirements during the reconstruction, deep learning (DL) can be used: Reconstruction is now performed by forward passing the signal (or signals) through a (regression) network, which is able to predict the $T_1$ and $T_2$ relaxation times for the input. Proposed approaches vary from Fully Connected Neural Networks (FCNs)~\cite{cohen2018mr}, Convolutional Neural Networks (CNNs)~\cite{fang2017quantification,hoppedeep,hoppe2017deep} and other architectures, e.g. incorporating an U-Net~\cite{fang2018deep}. However, also state-of-the-art DL solutions have their drawbacks: While FCNs are known to tend to overfit because of the huge number of parameters, CNNs are not optimally suited for time-resolved tasks. To overcome these limitations, we propose Recurrent Neural Networks (RNNs) for this reconstruction task due to their capabilities to capture the time dependency in the signal better than e.g. CNNs. We evaluate our approach using in-vivo data from multiple brain slices and several volunteers and investigate with an extensive evaluation following aspects: (1) the superior performance of RNNs over CNNs, (2) complex-valued input signal data instead of magnitude data as in some previous approaches (e.g.~\cite{cohen2018mr,hoppedeep}) and (3) spatially connected signal patches instead of one signal for the input layer in combination with a novel quantile filtering layer prior to the output layer. We expect small, spatially connected patches to have the same type of tissue and therefore the same quantitative parameters. The knowledge of spatial neighbors was shown to help the reconstruction accuracy by e.g.~\cite{fang2018deep}, but they used the whole image as input. To be able to train their network, all signals have to be compressed and possibly important information may be lost in the signals. Our approach uses smaller, not compressed patches of spatially connected signals (cf.~Fig.\ref{fig:overview}). To the best of our knowledge, RNNs for MRF were only investigated using signals from a synthetic dataset and without the consideration of spatial neighbor signals~\cite{oksuz2019magnetic}.

\section{Methods}
\subsection{Recurrent Neural Networks}\label{sec:rnns}
\paragraph{General architectures:} We devise a regression RNN to solve the MRF reconstruction task: From the input (one or more time signals), the network predicts the quantitative relaxation parameters for this signal. For the development of the networks, we use the well-known Long Short-Term-Memory (LSTM) layers~\cite{hochreiter1997long}. In order to keep the sequence in a moderate size, we reshaped the signals of length $n=3,000$ data points into 30 even sized parts. Thus, every sequence element consists of 100 complex-valued (flatterned to 200 values from the real and imaginary parts, respectively) or magnitude data points and is used in front of the LSTM layer as the first layer of our RNNs. This reshaping reduces the risk of vanishing or exploding gradient problems during the training~\cite{pascanu2013difficulty}. One LSTM layer is followed by the Rectified Linear Unit (ReLU) activation and a batch normalization (BN). Afterwards, we use 4 fully connected layers, each followed by a ReLU activation and a BN layer (each operating on either the magnitude or on the real and imaginery data points separately), to execute the regression. 
\paragraph{Quantile layer:} To cope with signal outliers due to undersampling or noise during the acquisition, we propose to combine the RNN architecture with a quantile layer as the last layer prior to the output. Inspired by work from Schirrmacher et al. in~\cite{schirrmacher2018temporal}, we use small $3 \times 3$ patches of signals, which are locally connected for the input layer. Thus, the input for one regression is increased by a factor of 9 compared to networks with one signal as input. For the output, we compute the $0.5$ quantile of all predictions from this neighborhood. The quantile operation $q()$ can be reformulated as $q(f) = \boldsymbol{Q}f$, where $\boldsymbol{Q}$ denotes a sparse matrix which stores the position of the quantile. In the backward pass, the gradient w.r.t. the input is simply the transposed matrix $\boldsymbol{Q}^{T}$. We expect the signals from small patches to belong to similiar or same parameters as they originate from same or similiar tissue type. The quantile layer enables a pooling operation that is more robust to noise compared to common pooling operations such as maximum or average pooling. To the best of our knowledge, we are the first to incorporate this operation as a network layer.

\subsection{Training and Evaluation}
All our models are trained based on the mean squared error (MSE) loss and optimized using ADAM. We evaluate all models by measuring the difference between the predicted and the ground-truth $T_1$ and $T_2$ relaxation times, computed as the relative mean error and the appropriate standard deviation. Data is split into disjunct training, validation and test sets. The validation set is used to select the best model from all training epochs, the test set for testing a model on unknown data afterwards. 

\section{Experiments and Results}
\subsection{Data sets}
\paragraph{Data acquisition:} All data sets for our experiments were measured as axial brain slices in 8 volunteers (4 male, 4 female, 43$\pm$15 years) on a MAGNETOM Skyra 3T MR scanner (Siemens Healthcare, Erlangen, Germany) using a prototype sequence based on Fast Imaging with Steady State Precession with spiral readouts~\cite{jiang2015mr} and following sequence parameters: Field-of-View: 300 mm, resolution: $1.17 \times 1.17 \times 5.0$ mm$^3$, variable TR (12-15 ms), FA (5-74$\degree$), number of repetitions: 3,000, undersampling factor: 48. From 2 volunteers, 2 different slices were available, from 6 volunteers, 4 slices were available each. All slices were measured at different positions and points in time to reduce possible correlations between slices from one volunteer.
\paragraph{Ground-truth data:} In order to create accurate ground-truth data for our DL experiments, we used a fine resolved dictionary containing overall 691,497 possible parameter combinations with $T_1$ in the range of 10 to 4,500 ms and $T_2$ of 2 to 3,000 ms, respectively. To be able to reconstruct the relaxation maps in a reasonable time and to reduce the memory requirements, the dictionary and measured signals were compressed to 50 main components in the time domain using SVD prior to the template matching.

\subsection{Experiments for finding architectural settings}
\paragraph{Experimental setup:} We ran three specific types of experiments to investigate following issues:
\begin{enumerate}
	\item Performance of networks using magnitude input signals $S_m\in \mathbb{R}$ vs. complex-valued input signals $S_c\in \mathbb{C}$. For this, we compared the CNN (architectural details see Section~\ref{sec:cnn}) and RNN models with $1\times1$ $S_{m}$ and $S_{c}$.
	\item Performance of networks using CNN vs. RNN models (both with a comparable number of learnable parameters). For this, we compared the CNN and RNN models with $1\times1$ input signals $S_{c}$.
	\item Performance of networks using $1\times1$ input signals $S_{c}$ vs. $3\times 3$ input signals $S_{c}$ in combination with a  $0.5$ quantile layer prior to the output. For this, we compared RNN models with and without a quantile layer.
\end{enumerate}
\paragraph{Data splitting:} As only a limited amount of data sets (overall 12 slices from 4 volunteers) was available for our extensive experiments, we first used all slices from these 4 volunteers randomly separated into training, validation and test sets (8, 2 and 2 slices, respectively). We then used additional 16 slices from another 4 volunteers (again randomly separated) for experiments with our best fitted model (19 slices for training, 7 for validation, 2 for testing).

\subsection{Comparison with other DL architecture}\label{sec:cnn}
We used the CNN model with overall 4 convolutional and 4 fully connected layers with ReLU activations and average pooling in~\cite{hoppedeep} to compare our approach with another DL based MRF reconstruction framework. We extended this baseline model with BN layers after each convolutional and fully connected layer. 

\subsection{Results}
Results can be found in Table~\ref{tab:results} (validation loss from the best epoch) and in Figure~\ref{fig:results} (parameter maps on the same test set from all models).

{\def\arraystretch{0.6}\tabcolsep=10.5pt
	\begin{table}
		\caption{Validation losses across different experiments. Best results are marked in bold. The validation loss is measured as $\sqrt{\text{MSE}}$ over $T_1$ and $T_2$ values. CNN$_1$: CNN model with $1\times1$ input signals, RNN$_1$: RNN model with $1\times1$ input signals, RNN$_3$: RNN model with $3\times3$ signal patch as input and quantile layer, RNN$_3^*$: the same as RNN$_3$, trained with the larger data set. Detailed information about the models see Sections~\ref{sec:rnns} and~\ref{sec:cnn}.}
		\label{tab:results}
		\begin{center}
			\begin{tabular}{ c  c c c  c}
				& \multicolumn{4}{c}{Validation loss [ms]}\\ \toprule
				Input signals & \multicolumn{4}{c}{Model} \\\toprule
				& CNN$_1$ & RNN$_1$ & RNN$_3$ & RNN$_3^*$ \\\cmidrule{2-5}
				$S_{m}\in\mathbb{R}$ & 636.96 &424.96 & - & - \\
				$S_{c}\in\mathbb{C}$ & 470.26 &269.20  &\textbf{221.52} & \textbf{195.34}\\\bottomrule
			\end{tabular}
		\end{center}
		
	\end{table}
}

\begin{figure}
	\centering
	\resizebox{0.8\textwidth}{!}{
		
		{\def\arraystretch{0.2}\tabcolsep=0.0pt 
			
			\begin{tabular}{cccc}
				\includegraphics[width=0.25\textwidth]{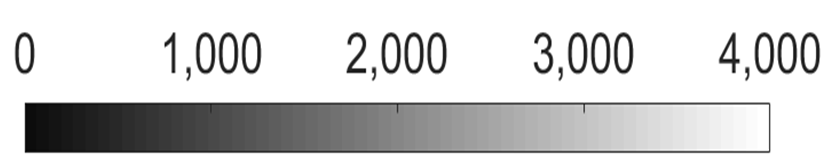}
				&\includegraphics[width=0.25\textwidth]{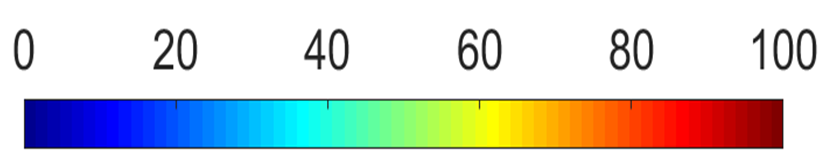}%
				&\includegraphics[width=0.25\textwidth]{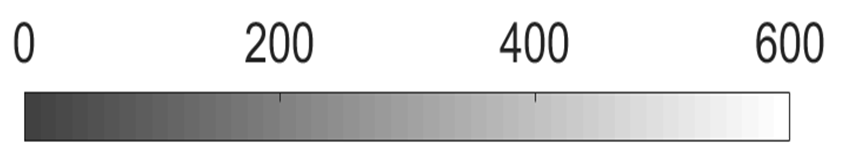}%
				&\includegraphics[width=0.25\textwidth]{plots/ErrorColorbarScaled.png}%
				\\
				\includegraphics[width=0.25\textwidth]{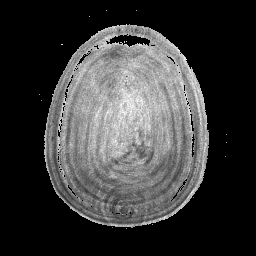}
				&\includegraphics[width=0.25\textwidth]{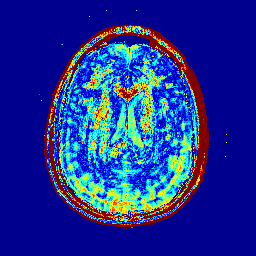}%
				&\includegraphics[width=0.25\textwidth]{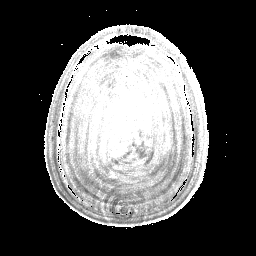}%
				&\includegraphics[width=0.25\textwidth]{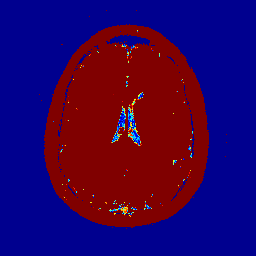}%
				\\
				\multicolumn{4}{c}{CNN with $1\times1$ $S_{m}$, RME\,$\pm$\,std.dev.[\%]: $T_1:70.6\pm122.4$, $T_2:1069.1\pm1363.2$} 
				\\
				\includegraphics[width=0.25\textwidth]{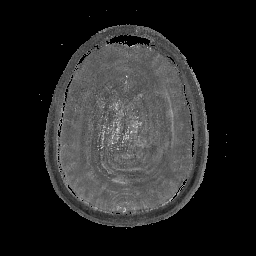}
				&\includegraphics[width=0.25\textwidth]{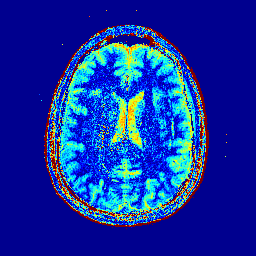}%
				&\includegraphics[width=0.25\textwidth]{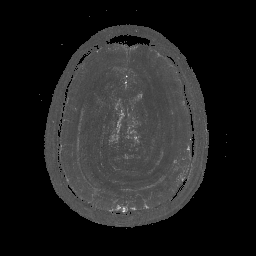}%
				&\includegraphics[width=0.25\textwidth]{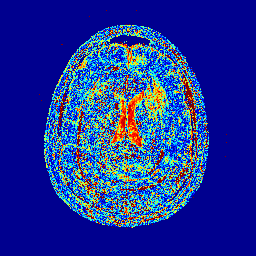}%
				\\
				\multicolumn{4}{c}{CNN with $1\times1$ $S_{c}$, RME\,$\pm$\,std.dev.[\%]: $T_1:43.8\pm81.4$, $T_2:48.1\pm112.8$}
				\\
				\includegraphics[width=0.25\textwidth]{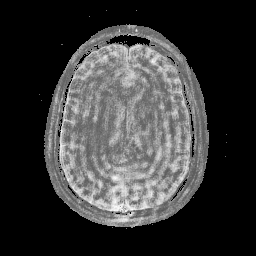}
				&\includegraphics[width=0.25\textwidth]{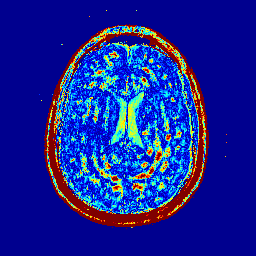}%
				&\includegraphics[width=0.25\textwidth]{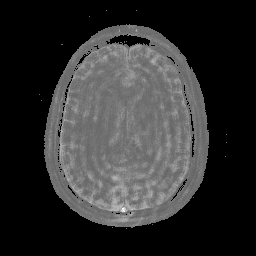}%
				&\includegraphics[width=0.25\textwidth]{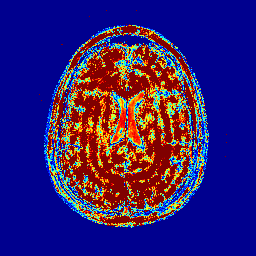}%
				\\
				\multicolumn{4}{c}{RNN with $1\times1$ $S_{m}$, RME\,$\pm$\,std.dev.[\%]: $T_1:64.3\pm108.7$, $T_2:108.3\pm156.9$ }
				\\
				\includegraphics[width=0.25\textwidth]{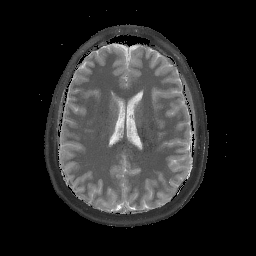}
				&\includegraphics[width=0.25\textwidth]{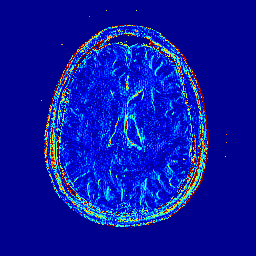}%
				&\includegraphics[width=0.25\textwidth]{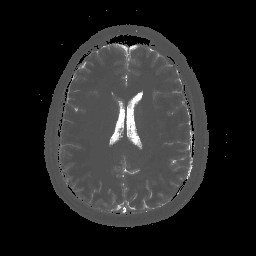}%
				&\includegraphics[width=0.25\textwidth]{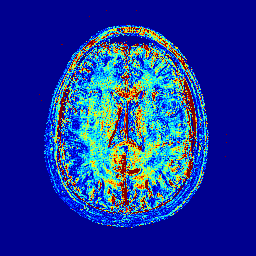}%
				\\
				\multicolumn{4}{c}{RNN with $1\times1$ $S_{c}$, RME\,$\pm$\,std.dev.[\%]: $T_1:23.5\pm50.7$, $T_2:55.0\pm217.8$ }
				\\
				\includegraphics[width=0.25\textwidth]{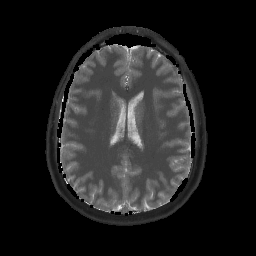}
				&\includegraphics[width=0.25\textwidth]{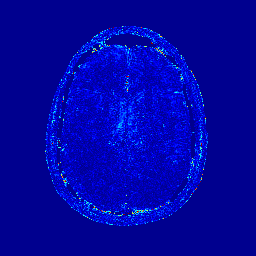}%
				&\includegraphics[width=0.25\textwidth]{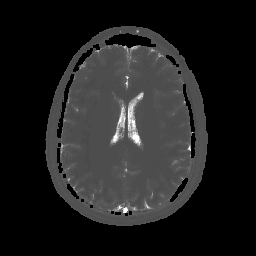}%
				&\includegraphics[width=0.25\textwidth]{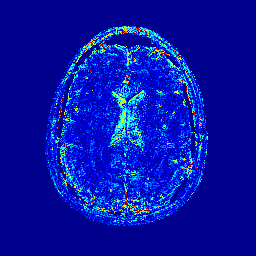}%
				\\
				\multicolumn{4}{c}{RNN with $3\times3$ $S_{c}$, RME\,$\pm$\,std.dev.[\%]: $T_1:13.6\pm25.3$, $T_2:23.9\pm66.8$ }
				\\
				\includegraphics[width=0.25\textwidth]{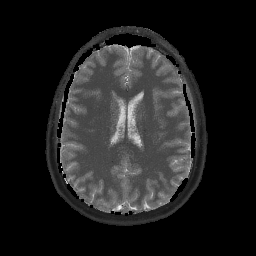}
				&\includegraphics[width=0.25\textwidth]{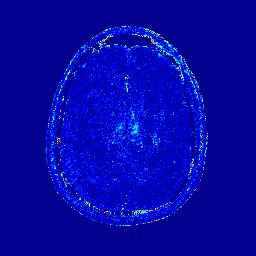}%
				&\includegraphics[width=0.25\textwidth]{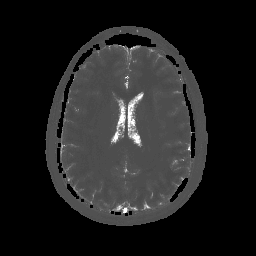}%
				&\includegraphics[width=0.25\textwidth]{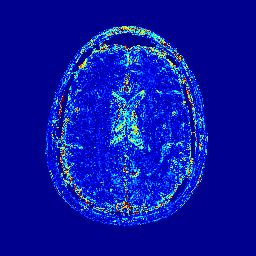}%
				\\
				\multicolumn{4}{c}{RNN with $3\times3$ $S_{c}$, RME\,$\pm$\,std.dev.[\%]: $T_1:14.9\pm27.2$, $T_2:26.7\pm94.9$}
			\end{tabular}%
			
	}}
	\caption{Predicted maps of one test data set from models using small data set (rows 1-5), or large data set (row 6). First column: $T_1$ maps. Second column: $T_1$ relative mean errors to the ground-truth. Third column: $T_2$ maps. Fourth column: $T_2$ relative mean errors to the ground-truth. For better visibility, all relative error maps were clipped at 100\,\%, the background of all $T_1$ and $T_2$ maps was set to -200 and they were windowed equally for fair comparison (0 - 4,000 ms and 0 - 600 ms, respectively). RME: Relative mean error, std.dev.: standard deviation.} 
	\label{fig:results}	
\end{figure}

\section{Discussion}
In summary, the main observation from our results is the clear improvement of the performance using our proposed RNN model in combination with complex-valued input signals and the quantile layer in comparison to all other tested models.  
\paragraph{Magnitude vs. complex-valued signal inputs:} We first compare our models trained with $S_{m}$ and $S_{c}$ inputs. The utilization of both components of the complex-valued signals, instead of only computing the magnitudes for the input layers of the networks, is an essential factor for the performance. A clear reduction of the errors is achieved using $S_{c}$ for both approaches (CNN: more than 62\,\%, RNN: more than 50\,\%). Comparing the visual results of e.g. the same RNN model using $S_{m}$ and $S_{c}$ (cf. rows 3, 4 in Fig.~\ref{fig:results}), the complex version clearly yields reduced relative mean errors and improved parameter maps without being corrupted by the heavy ringing artifacts which appear with the $S_{m}$ inputs.
\paragraph{CNN vs. RNN:} A clear improvement is also achieved using a RNN instead of a CNN model with a reduction of the errors up to 53\,\%. Independent of the input signal types, the CNN model is not able to reconstruct meaningful parameter maps showing soft tissue contrast. In comparison, the RNN model is capable of reconstructing high detail parameter maps, showing the better capability of the RNN for processing time-dependent signals. Nevertheless, this holds only for the RNN using $S_c$, since the RNN using $S_m$ is still corrupted by the ringing artifacts. 
\paragraph{Quantile layer:} Our results show additionally, that a quantile layer furthermore improves the performance (cf. rows 4, 5 in Fig.~\ref{fig:results}), reducing the errors by 57\,\% and 43\,\% for $T_1$ and $T_2$, respectively, in comparison to a RNN without quantile layer. The influence of the quantile layer is particularly evident at transitions between different tissue types in the parameter map. With the help of the quantile layer, the errors at the edges can be enormously reduced, as the 0.5 quantile layer acts as an edge-preserving denoising filter (cf. the relative error maps in rows 4, 5 in Fig.~\ref{fig:results}).
\paragraph{Challenges and limitations:} Our experiments show the  improved performance step-by-step, that increases from (1) magnitude to complex-valued input signals, (2) from a CNN to a RNN model and (3) from a RNN without a quantile layer to a RNN with a quantile layer. Even though we use a limited amount of data, our results are a strong indication, that our model is able to generalize. Using our best RNN model and training it with slightly more data already decreased the error (cf. Table~\ref{tab:results}), which encourages this assumption. One further step, however, is the evaluation of our proposed approach using data splits with completely unseen volunteer data sets in the validation or test data when more data is available (preliminary experiments in this direction are attached in the Supplementary Material). Moreover, we used a very fine-resolved dictionary for the ground-truth data. While this is crucial for accurate ground-truth data, this further increases the amount of training data that is necessary to fully imprint the complex mapping into the network. In comparison to other MRF DL approaches (e.g. the MRF-EPI sequence in~\cite{cohen2018mr}), we used signals from a very strongly undersampled acquisition (undersampling factor: 48), which leads to very noisy and corrupted signals compared to simulated dictionary signals. As shown by Hoppe et al. in~\cite{hoppedeep,hoppe2017deep}, fully sampled dictionary signals can be easily learned by simple CNN models. However, undersampled in-vivo data are more challenging to reconstruct with the MRF DL method, thus a more complex model is required.

\section{Conclusion}
We proposed a regression RNN for MRF reconstruction. Our architecture combines a model used to deal with time-dependent complex-valued input signals incorporated as a LSTM layer with a novel quantile layer to deal with signal outliers, which are very common due to the strong undersampling during the acquisition. We evaluated our approach in a proof-of-concept study with various experiments and showed, that our model outperforms other DL models like CNNs or RNNs without the additional quantile layer, reducing the errors by more than 80\,\%. One limitation of our study is the restricted amount of training data, which will be addressed in future work. Furthermore, another future step will be a deeper comparison of the different architectures and their features which can help to improve the interpretability of the networks. In addition, the incorporation of known operations based on the imaging physics within the networks as described in~\cite{maier2019learning} can help to reduce the complexity and improve the performance at the same time. This also will be investigated for our application.

%
%
%
\bibliographystyle{splncs04}
	\bibliography{main}

	\appendix
	\title{Supplementary Material}	
	\author{}
	\institute{}
	\authorrunning{E.~Hoppe et al.}
	\titlerunning{RinQ Fingerprinting: Supplementary Material}
	\maketitle
	\pagenumbering{gobble}
	
\section{Dictionary parameters for ground-truth data}
{\def\arraystretch{0.6}\tabcolsep=10.5pt
	\begin{table}
		\caption{Parameter steps used for the dictionary for generating ground-truth parameter maps.}
		\begin{center}
			\begin{tabular}{ c  c   }
				$T_1$ start:step:stop[ms] & $T_2$ start:step:stop[ms]\\\toprule			
				10:10:90 & 2:2:98  \\
				100:20:1000 & 100:5:150   \\
				1040:40:2000 & 160:10:300  \\
				2050:100:4500& 350:50:800  \\
				& 900:100:1600  \\
				& 1800:200:3000 \\\bottomrule  
			\end{tabular}
		\end{center}
		
	\end{table}
}
\clearpage

\section{Deep learning architectures}
\subsection{Overview}
\begin{figure}
	\centering
	\includegraphics[width=\textwidth]{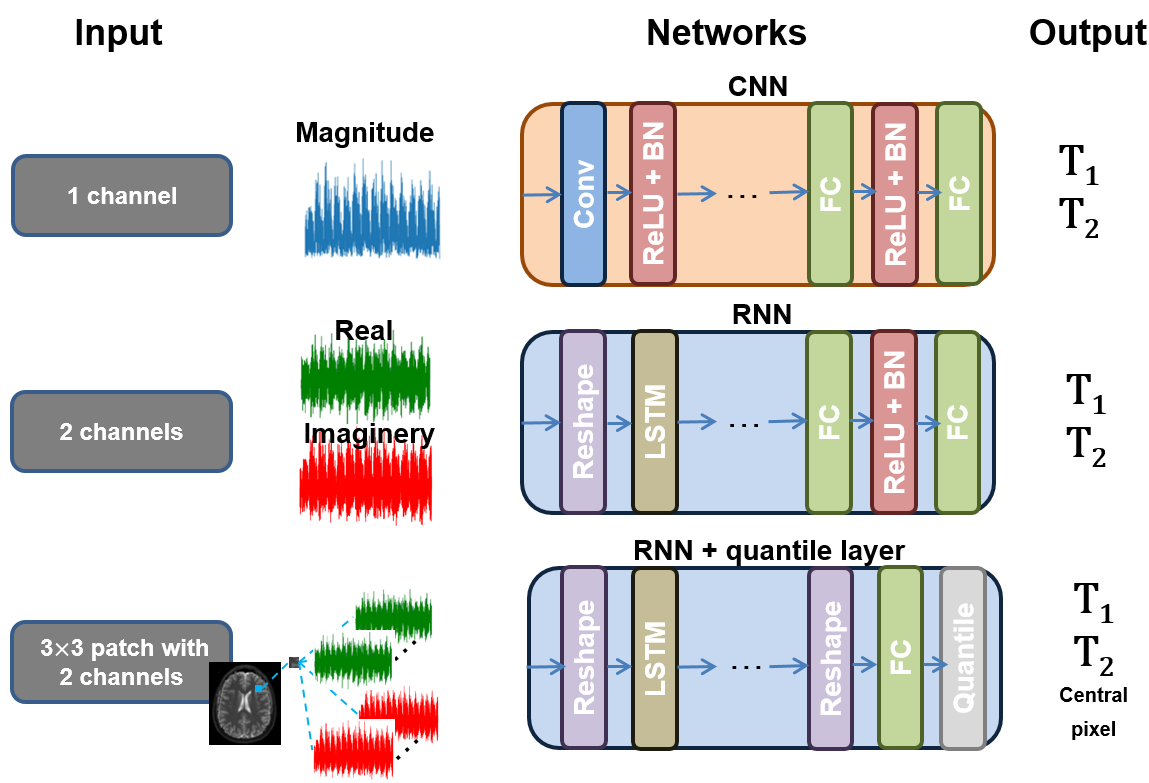}
	\caption{Overview over the used models and input signal types in our work (not all layers within the networks are displayed). We used models with magnitude (upper model) and complex-valued input signals (middle and lower models). Furthermore, we investigated Convolutional Neural Networks (CNNs, upper model) and different Recurrent Neural Networks (RNNs, with and without a quantile layer prior to the output layer, the middle and the lower model, respectively). Conv: Convolutional layer, ReLU: Rectified Linear Unit, BN: Batch normalization, FC: Fully Connected layer, LSTM: Long Short-Term Memory layer.}
\end{figure}
\clearpage

\subsection{Recurrent Neural Networks}
{\def\arraystretch{0.6}\tabcolsep=10.5pt
	\begin{table}
		\caption{Details of different layers used for our RNN model processing signals from one pixel with one channel (magnitudes) and two channels (real and imaginery parts from the complex numbers) as input. After every layer (except for the Input, Reshape and Flatten) a Rectified Linear Unit activation is applied. LSTM: Long Short-Term Memory layer, FC: Fully Connected layer. Number of parameters [millions]: 7.7 (RNN with one channel signals), 8.1 (RNN with two channels signals).}
		\begin{center}
			\begin{tabular}{ c  c c}
				Layer & \multicolumn{2}{c}{Output shape} \\
				& One channel signals & Two channels signals \\\toprule
				Input &	(3,000, 1) & (3,000, 2)\\
				Reshape & (30, 100)& (30, 200)\\
				LSTM & (30, 1,000)&(30, 1,000)\\
				FC$_1$ & (30, 500)&(30, 500)\\
				FC$_2$ & (30, 250)&(30, 250)\\
				Flatten & (7500)& (7500)\\
				FC$_3$ & (360)&(360)\\
				FC$_4$ & (2) &(2)\\\bottomrule
			\end{tabular}
		\end{center}
		
	\end{table}
}
{\def\arraystretch{0.6}\tabcolsep=10.5pt
	\begin{table}
		\caption{Details of different layers used for our RNN model processing signals from 3$\times$3 patches and two channels (real and imaginery parts from the complex numbers) as input. After every layer (except for the Input, Reshape and Flatten) a Rectified Linear Unit activation is applied. LSTM: Long Short-Term Memory layer, FC: Fully Connected layer. Number of parameters [millions]: 7.7.}
		\begin{center}
			\begin{tabular}{ c  c }
				Layer & Output shape \\\toprule
				Input &	(3, 3, 3,000, 2)\\
				Reshape$_1$ & (30, 1800)\\
				LSTM & (30, 500)\\
				FC$_1$ & (30, 500)\\
				FC$_2$ & (30, 250)\\
				Flatten & (7500)\\
				FC$_3$ & (360)\\
				Reshape$_2$ & (3, 3, 40) \\
				FC$_4$ & (3, 3, 2)\\		
				Quantile & (2) \\\bottomrule
			\end{tabular}
		\end{center}
		
	\end{table}
}
\clearpage
\subsection{Convolutional Neural Networks}
{\def\arraystretch{0.6}\tabcolsep=10.5pt
	\begin{table}
		\caption{Details of different layers used for our CNN model processing signals from one pixel as input. After every layer (except for the Input, AvgPool and Flatten) a Rectified Linear Unit activation is applied. Conv: Convolutional layer, BN: Batch normalization, AvgPool: Average Pooling, FC: Fully Connected layer, K: Kernel size, S: Stride size. Number of parameters [millions]: 6.3.}
		\begin{center}
			\begin{tabular}{ c  c c c}
				Layer & Kernel sizes & \multicolumn{2}{c}{Output shape} \\
				& & One channel signals & Two channels signals \\\toprule
				Input& & (3,000, 1)& (3,000, 2)\\		
				Conv$_1$ + BN&K: 15$\times$1, S: 5 &(598, 30)& (598, 30)\\	
				Conv$_2$ + BN& K: 10$\times$1, S: 3&(197, 60)&(197, 60) \\	
				Conv$_3$ + BN&K: 5$\times$1, S: 2 &(97, 120)& (97, 120)\\
				Conv$_4$ + BN &K: 3$\times$1, S: 2&(48, 240)& (48, 240)\\
				AvgPool &K: 3, S: 2&(23, 240)&(23, 240)\\
				Flatten &&(5520)&(5520)\\
				FC$_1$ + BN&&(1,000)&(1,000)\\
				FC$_2$ + BN&&(500)&(500)\\
				FC$_3$ + BN&&(300)&(300)\\
				FC$_4$&&(2)&(2)\\\bottomrule  
			\end{tabular}
		\end{center}
		
	\end{table}
}

\subsection{Training parameters}
{\def\arraystretch{0.6}\tabcolsep=10.5pt
	\begin{table}
		\caption{Training parameters used for all our models and experiments. Other training parameters for our optimizer ADAM are: $\beta_1=0.9$, $\beta_2=0.999$. CNN$_1$: CNN model with $1\times1$ input signals, RNN$_1$: RNN model with $1\times1$ input signals, RNN$_3$: RNN model with $3\times3$ signal patch as input and quantile layer, RNN$_3^*$: the same as RNN$_3$, trained with the larger data set. $S_m\in \mathbb{R}$: magnitude input signals, $S_c\in \mathbb{C}$: complex-valued input signals.}
		\begin{center}
			\begin{tabular}{ c c c c}
				\multicolumn{2}{c}{Model} & Learning rate & Batch size \\\toprule	
				\multirow{2}{*}{$S_{m}\in \mathbb{R}$}& CNN$_1$ & \multirow{2}{*}{$5\times 10^{-5}$} & \multirow{2}{*}{128}\\	
				&RNN$_1$ & & \\\midrule 
				\multirow{4}{*}{$S_{c}\in\mathbb{C}$}& CNN$_1$ & $5\times10^{-5}$ & \multirow{2}{*}{128}\\
				& RNN$_1$ &$10^{-3}$ &  \\
				& RNN$_3$ & \multirow{2}{*}{$10^{-4}$}& \multirow{2}{*}{32}\\
				& RNN$_3^*$&&  \\\bottomrule
			\end{tabular}
		\end{center}
		
	\end{table}
}
\clearpage

\section{Leave-one-out splits: Preliminary results}

\begin{figure}
	\centering
	\resizebox{\textwidth}{!}{
		{\def\arraystretch{0.2}\tabcolsep=0.0pt 
			
			\begin{tabular}{cccc}
				\includegraphics[width=0.25\textwidth]{plots/T1ColorbarScaled.png}
				&\includegraphics[width=0.25\textwidth]{plots/ErrorColorbarScaled.png}%
				&\includegraphics[width=0.25\textwidth]{plots/T2ColorbarScaled.png}%
				&\includegraphics[width=0.25\textwidth]{plots/ErrorColorbarScaled.png}%
				\\
				\includegraphics[width=0.25\textwidth]{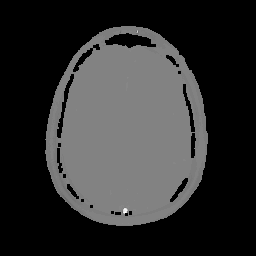}
				&\includegraphics[width=0.25\textwidth]{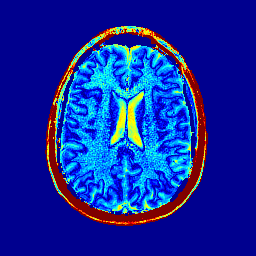}%
				&\includegraphics[width=0.25\textwidth]{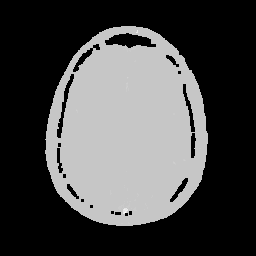}%
				&\includegraphics[width=0.25\textwidth]{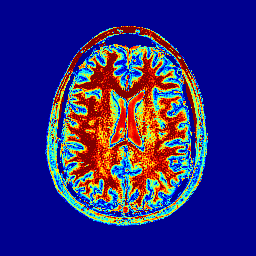}%
				\\
				\includegraphics[width=0.25\textwidth]{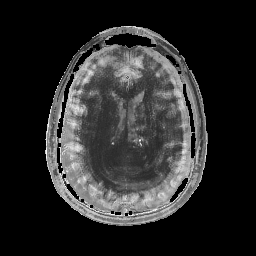}
				&\includegraphics[width=0.25\textwidth]{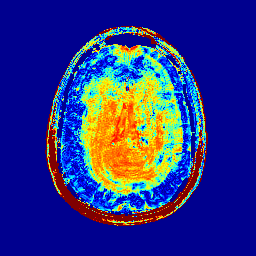}%
				&\includegraphics[width=0.25\textwidth]{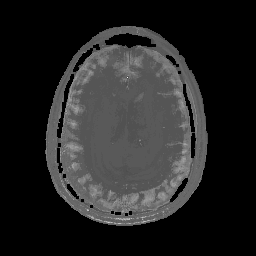}%
				&\includegraphics[width=0.25\textwidth]{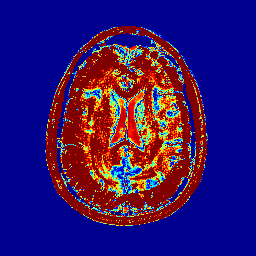}%
				
			\end{tabular}%
			
	}} 
	\caption{Predicted maps of one test data set from models using leave-one-out data separation with the small data set (overall 12 slices from 4 volunteers, row  1), or using leave-one-out data separation with the extended data set (overall 28 slices from 8 volunteers, row 2). First column: $T_1$ maps. Second column: $T_1$ relative mean errors to the ground-truth. Third column: $T_2$ maps. Fourth column: $T_2$ relative mean errors to the ground-truth. For better visibility, all relative error maps were clipped at 100\,\%, the background of all $T_1$ and $T_2$ maps was set to -200 and they were windowed equally for fair comparison (0 - 4,000 ms and 0 - 600 ms, respectively). Every data set was separated using slices from one previously unseen volunteer for the validation and the test processes, respectively (resulting in 2 (small data set) or 6 (extended data set) volunteers for training, 1 volunteer for validation and 1 volunteer for testing). The homogeneous areas in the reconstructed parameter maps from the training with only 2 volunteer data sets (row 1) show, that these data sets are not sufficient for the model to be able to generalize. The same experiment with an extended amount of data sets to 6 volunteers in the training phase already shows an enormous increase of the performance, as tissue details can be recognized in the reconstructed parameter maps (row 2). With this example, we would like to emphasize that for the present case even 6 volunteer records are not nearly enough training data, but the improvements from 2 to 6 volunteer data sets are tremendous. The results of the leave-one-out experiments can be seen as the lower limit for future results with more volunteer training data.}

\end{figure}

\end{document}